\documentclass[aps,prb,twocolumn,showpacs,superscriptaddress,amssymb]{revtex4}
\usepackage{graphicx}
\usepackage{amsmath}
\usepackage{color}

\newcommand{\etal}{\textit{et al.}}
\newcommand{\abin}{\textit{ab initio}}

\begin{document}

\title{Can hydrogen be stored inside carbon nanotubes
       under pressure in gigapascal range?}

\author{X. H. Zhang}
\affiliation{Department of Physics, Fudan University,
Shanghai 200433, China}
\affiliation{Department of Chemistry, Chinese University of Hong Kong,
Shatin, Hong Kong, China}

\author{X. G. Gong}
\affiliation{Department of Physics, Fudan University,
Shanghai 200433, China}

\author{Z. F. Liu}
\affiliation{Department of Chemistry, Chinese University of Hong Kong,
Shatin, Hong Kong, China}

\date{\today}


\begin{abstract}
By using a newly fitted multi-parameter potential to describe the
van der Waals interaction between carbon and molecular hydrogen,
we study the hydrogen storage inside carbon nanotubes (CNT's) under
pressure in gigapascal range.
Comparing with the results of graphite, we find that the shape change
of the nanotubes (the curvature effect) provides a different storage
mechanism for hydrogen.
The negative free energy change for hydrogen storage inside CNT's
makes it possible to use CNT's as the nanocontainer
[Carbon {\bf 45}, 315 (2007)].
\end{abstract}

\maketitle

\section{Introduction}

Technologies using hydrogen as an energy source are being developed
rapidly, among which the hydrogen storage in carbon materials is one
major aspect. \cite{schlapbach.l:2001}
Dillon {\etal} \cite{dillon.ac:1997} are the first that
suggested the possibility of achieving very high hydrogen storage
capacity by using single-wall carbon nanotubes (SWCNT's).
By experiments using high-purity SWCNT's, hydrogen storage capacity
of about 8 wt\% at 80 K and 120 atm was reported by Ye {\etal},
\cite{ye.y:1999} and about 4.2 wt\% at room temperature and 100 atm
by Liu {\etal}. \cite{liu.c:1999}
On the other hand, molecular dynamics simulations and the Monte Carlo
method were used to study the hydrogen adsorption.
\cite{wang.q:1999,darkrim.f:1998}
However, there are still many open questions to be resolved, for example,
whether hydrogen is chemisorbed or physisorbed in carbon materials and
whether the intratube space of SWCNT's is more important than
the intertube space.

The report by Ma {\etal} \cite{ma.y:2001} that chemisorption is
achieved by collision of H atom with energy of 1-3 eV,
and the report by Chan {\etal} \cite{chan.sp:2001} on the breaking
of the H-H bond between carbon nanotubes under high pressure show
the possibility of the chemisorption.
Very recently, an experimental study should be a breakthrough for the
chemisorption, \cite{nikitin.a:2005} that carbon nanotube films are found
to be hydrogenated and the C-H bonds can be broken by heating to
600$^\circ$C.

However, for the much more widely studied physisorption, there are very few
exciting results.
In 2004, Chan {\etal} \cite{chan.sp:2004} found by {\abin}
calculation that H$_2$ molecules can be well confined between graphene
sheets under pressure in the gigapascal range due to the negative free
energy change $\Delta G$.
This method is much better than many of the other simulations, because
in those studies either the external pressure is not considered or the
interaction between carbon atoms and H$_2$ molecules are not well described.
For example, Ma and Xia {\etal} \cite{ma.y:2001a,ma.y:2002,xia.y:2003}
studied the energy between H$_2$ molecules and the wall of the SWCNT
and found helical structures of hydrogen inside the tube.
However, we don't know whether the H$_2$ molecules want to stay inside
the tube or not, because the tube in their studies is encapsulated.
Furthermore their potentials describing the C-H$_2$ and H$_2$-H$_2$
interactions are both Lennard-Jones (LJ) like, which are not accurate
under high pressure \cite{silvera.if:1978,sun.dy:2006}.

In this study, by using molecular dynamics simulations with a more
accurate multi-parameter potential for C-H$_2$ van der Waals (vdW) interaction,
\cite{sun.dy:2006}
we investigate the hydrogen storage inside the graphene sheets and
carbon nanotubes under pressure.
We find that the carbon nanotubes have a different storage mechanism
from the graphite.
For the graphite system, one must separate the graphene sheets
by overcoming a large binding energy to store the H$_2$ molecules.
But in the nanotubes, such binding energy is easy to overcome because
the storage of hydrogen relaxes the tube from fully collapsed shape
(thin ellipse) to fat ellipse and provides more negative contribution
to the total energy change.
This is the main reason that hydrogen storage inside nanotubes
might has a negative free energy change which is the key criterion
for the storage.

Furthermore, our research also shows the possibility of CNT-based
nanocontainer, \cite{ye.x:2007} in which the hydrogen can be filled
inside the container under high pressure and can be locked inside
the tube due to molecular valves located at the two ends of the tube.

\section{Methodology}

\textsl{Energy changes --}
To calculate the free energy change
we simulate pure molecular hydrogen (H$_2$), pure carbon system (C), and
hydrogen-intercalated carbon system (C+H$_2$), respectively.
Thus the free energy change can be written as
\begin{align}
\Delta G &=\Delta H-T\Delta S=\Delta E+p\Delta V-T\Delta S\nonumber\\
         &=\Delta E+p\Delta V-T(S(\mbox{C+H$_2$})
           -[S(\mbox{C})+S(\mbox{H$_2$})])\nonumber\\
   &\approx\Delta E+p\Delta V+TS(\mbox{H$_2$}).
\label{eqn:deltaG}
\end{align}
Here we consider the entropy difference between pure and
hydrogen-intercalated carbon system as zero, because under high pressure,
the H$_2$ molecules inside carbon systems are well confined to form
an ordered structure.

In order to study the hydrogen storage mechanism, we also divide
the energy into four parts,
\begin{align}
\mbox{(C+H$_2$):} & \,\, E=E_{\mathrm{CC}}^1+E_{\mathrm{CC}}^2
  +E_{\mathrm{H}_2-\mathrm{H}_2}+E_{\mathrm{C-H}_2}, \\
\mbox{(C):}       & \,\, E=E_{\mathrm{CC}}^1+E_{\mathrm{CC}}^2, \\
\mbox{(H$_2$):}   & \,\, E=E_{\mathrm{H}_2-\mathrm{H}_2},
\end{align}
where $E_{\mathrm{CC}}^1$ and $E_{\mathrm{CC}}^2$ are the covalent
and the vdW energy between carbon atoms respectively,
$E_{\mathrm{H}_2-\mathrm{H}_2}$ is the energy between H$_2$ molecules,
and $E_{\mathrm{C-H}_2}$ is the energy between carbon
and molecular hydrogen.
Thus $\Delta E$ can be written in the form
\begin{equation}
\Delta E=\Delta E_{\mathrm{CC}}^1 + \Delta E_{\mathrm{CC}}^2
        +\Delta E_{\mathrm{H}_2-\mathrm{H}_2} + E_{\mathrm{C-H}_2}.
\label{eqn:deltaE}
\end{equation}

\textsl{Systems --}
The cubic simulation cell for pure hydrogen contains 2400 molecules.
For hydrogen storage inside graphite, we simulated one 1920-C graphite
and $N$-H$_2$-intercalated 960-C graphite where $N=480$, 960, and 1440,
respectively.
Periodic boundary conditions are used in all three directions
($x$, $y$, and $z$) for above simulation cells.

For the tubes, one 600-C (15,15) SWCNT and one 400-C (10,10) SWCNT
are chosen in our study.
Inside the tubes, we put different number of hydrogen, such as
150, 200, and 300 H$_2$ inside the (15,15) tube and 80, 100, and 150
H$_2$ inside the (10,10) tube.
The nanotubes are periodic along their axial direction ($z$).

\textsl{Simulation details --}
The carbon atoms are coupled to a heat bath by using the Berendsen velocity
scaling method, \cite{berendsen.hjc:1984} while all H$_2$ molecules
inside the carbon systems are free.
Simulations show the carbon atoms maintain at the set temperature 300 K
with reasonably small fluctuations and the H$_2$ molecules
can reach such temperature in picoseconds.
For pure hydrogen, the temperature is controlled by the same method.

For each direction with periodic boundary condition, the external pressure
is applied using the Berendsen algorithm, \cite{berendsen.hjc:1984}
while for the $xy$ plane of the carbon nanotubes, a constant-pressure
method for finite systems is used. \cite{sun.dy:2002}
We also use another method to apply the hydrostatic pressure on the tube
in the $xy$ plane, by introducing quite a large number of inert gas (He)
to transmit the pressure.
These two methods produce the same results.
However, the latter method is the real way to apply the pressure
in experiments.

The velocity verlet method is used to resolve the motion equation, with
the time step 1.0 fs.
After the system reaches the equilibrium, all data are got by averaging
in 50 ps.

\textsl{Empirical potentials --}
At temperature of 300 K and moderate pressure range,
H$_2$ molecules can be treated as structureless spherical particles,
and are modeled by the Silvera-Goldman potential. \cite{silvera.if:1978}
The C-C covalent bonds in graphite and carbon nanotubes are described
by using the Brenner potential, \cite{brenner.dw:1990} which has
been widely used for carbon systems.
For the C-C vdW interactions, a conventional treatment of LJ potential,
$V=C_{12}/r^{12}-C_{6}/r^6$, where $C_{12}=2.48\times10^4$
eV{\AA}$^{12}$ and $C_{6}=20$ eV{\AA}$^6$, is adopted.
By using such potential, at 300 K and zero pressure,
the binding energy for two graphene sheets is 30.7 meV/atom with
interplanar spacing of 3.30 {\AA}, which is in agreement
with both experimental and the density functional theory results
those range from 25 to 57 meV/atom with the same spacing.
\cite{chakarovakack.sd:2006}
For the C-H$_2$ vdW interactions, we use a more
accurate multi-parameter potential, \cite{sun.dy:2006}
\begin{align}
\phi(r)&=\exp(\alpha-\beta r-\gamma r^2)\nonumber\\
&-\left(\frac{C_6}{r^6}+\frac{C_8}{r^8}+\frac{C_{10}}{r^{10}}\right)
\times\exp\left[-\left(\frac{r_m}{r}-1\right)^2\right],
\end{align}
which is fitted from {\abin} results and can provide
a good description for high pressure regimes.
The parameters for this potential are listed
in Table~\ref{tab:parameter}.

\begin{table}[t]
\caption{Fitted parameters for the multi-parameter C-H$_2$ potential.}
\begin{ruledtabular}
\begin{tabular}{cr@{}lcr@{}l}
Parameter&\multicolumn{2}{c}{Value}&Parameter&\multicolumn{2}{c}{Value}\\
\hline
$\alpha$  &  1.55888 &              & $C_6$    &  15.978 & eV{\AA}$^6$\\
$\beta$   &        0 & {\AA}$^{-1}$ & $C_8$    &       0 & eV{\AA}$^8$\\
$\gamma$  & 0.642395 & {\AA}$^{-2}$ & $C_{10}$ & 1166.92 & eV{\AA}$^{10}$\\
$r_m$     &  6.21452 & {\AA}        &          &         &
\end{tabular}
\end{ruledtabular}
\label{tab:parameter}
\end{table}

\section{Hydrogen inside graphite}

\begin{figure*}
\centering
 \includegraphics[width=6in]{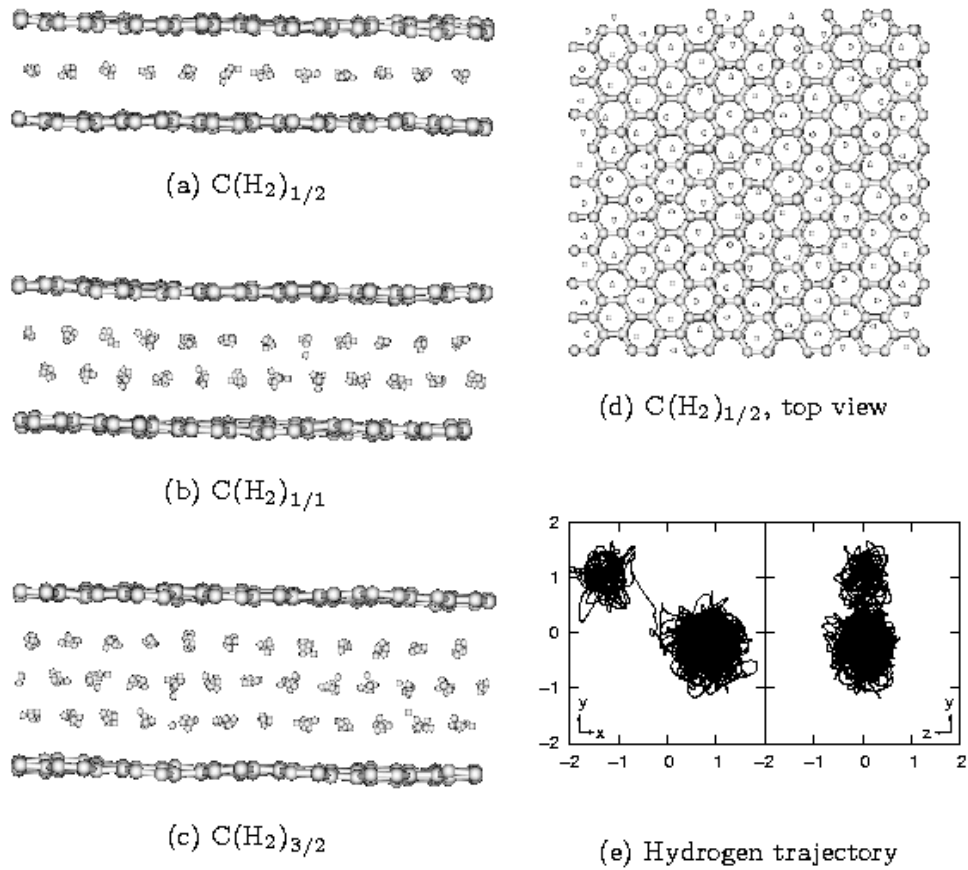}
\caption{Hydrogen structure between graphene sheets under 6.0 GPa
at 300 K. (a)-(c): different layers for different number of hydrogen;
(d): a top view for the (a) structure; (e): the trajectory of one
H$_2$ molecule for the (c) structure under 5.0 GPa. The trajectory
shows that the molecule hops from one confined position to another
and stays there for ever. Such hopping phenomenon just happens
under low pressure and at the time when the pressure increases.
The length unit for the trajectory is angstrom ({\AA}).}
\label{fig:graphite:structure}
\end{figure*}

\begin{figure}
\centering
\includegraphics[height=3.2in,angle=-90]{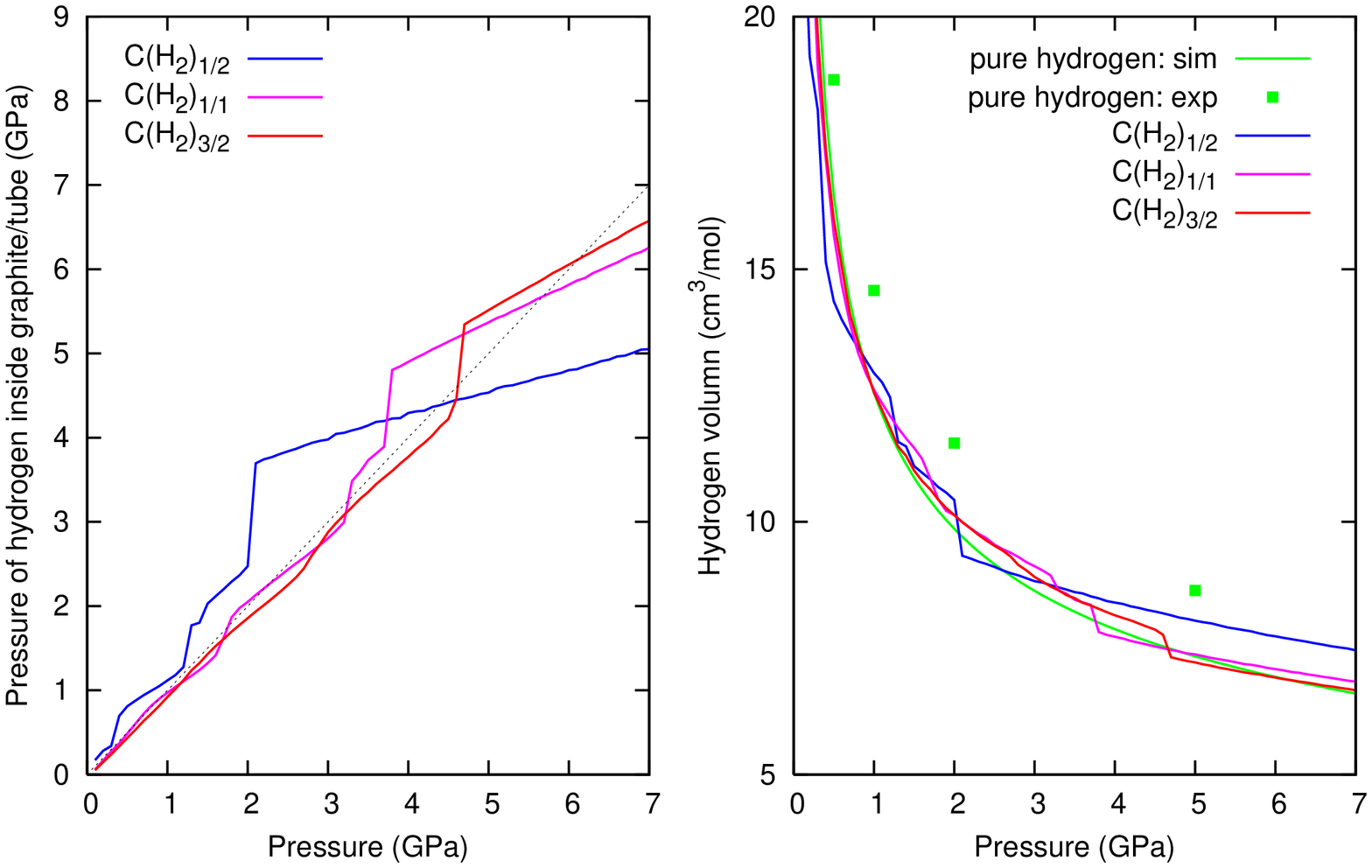}
\caption{The pressure (left) and the volume (right) of the hydrogen
between graphene sheets.
Green squares are taken from the experimental data \cite{hemmes.h:1986}
and the others are the results in our simulations.}
\label{fig:graphite:h2_rho}
\end{figure}

\textsl{Layered structure --}
We plot the hydrogen structure between two graphene sheets in
Figure~\ref{fig:graphite:structure}.
Layered structure is formed under high pressure with the number
of layer dependent on the molecule number.
The top view [Figure~\ref{fig:graphite:structure}(d)] shows
all the H$_2$ molecules are well confined between the hexagonal
rings of the graphene sheets, in agreement with the Sandwich structure
in Chan {\etal}'s study. \cite{chan.sp:2004}
Figure~\ref{fig:graphite:structure}(e) shows the trajectory
of one H$_2$ molecule in the three-layered structure under 5.0 GPa.
The molecule hops from one confined position to another in about
10 ps after the external pressure increases, and stays for ever.
Different from the liquid state under the same pressure,
the hydrogen is a solid because such hopping phenomenon
is very few and can not be found any more under higher pressure.
This is interesting because at room temperature the pure hydrogen
becomes solid under a higher pressure about 8.9 GPa. \cite{hemmes.h:1986}

Simulations also show the hydrogen inside the graphite reaches
this layered structure through two-step condensations.
For example, in the graphite system with the number ratio of
C:H$_2$=2:1, the gas hydrogen undergoes the first condensation
to form a two-layered structure under pressure about 0.5 GPa.
With increasing the pressure, the hydrogen becomes more liquid-like.
The second condensation happens under $\sim$ 2.0 GPa
to form a one-layered solid-like structure.
For the number ratio of 1:1 and 2:3, the pressures for the
second condensation are 3.7 and 4.6 GPa, respectively.

In Figure~\ref{fig:graphite:h2_rho} we show the hydrogen volume (right)
and the pressure of hydrogen along the $xy$ plane (left).

\begin{table*}[t]
\caption{Energy changes per hydrogen molecule for pressures
under which hydrogen is fully confined between
graphene sheets.
All three simulations for different carbon-hydrogen
number ratios show positive free energy change $\Delta G$,
which becomes more positive with increasing
the number of hydrogen.
Specially for the number ratio of C:H$_2$=2:1, $p\Delta V$, $T\Delta S$,
and the contributions to the total energy change $\Delta E$
are provided.}
\begin{ruledtabular}
\begin{tabular}{ccccccccccc}
Pressure & \multicolumn{3}{c}{$\Delta G$ for C(H$_2$)$_x$}
         & \multicolumn{5}{c}{$\Delta E$ for C(H$_2$)$_{1/2}$}
         & $p\Delta V$ for & \\
\cline{2-4}\cline{5-9}
(GPa)    & $x=1/2$ & $x=1/1$ & $x=3/2$
         & $\Delta E_{\mathrm{CC}}^1$
         & $\Delta E_{\mathrm{CC}}^2$
         & $\Delta E_{\mathrm{H}_2-\mathrm{H}_2}$
         & $E_{\mathrm{C-H}_2}$
         & $\Delta E$
         & C(H$_2$)$_{1/2}$ & $T\Delta S$\\
\hline
4.5      & 0.0939 & 0.1134 & 0.1338 & 0.0006 & 0.1026 &-0.0279 &-0.0965 &-0.0211 &-0.0137 & 0.1287\\
5.0      & 0.0820 & 0.1050 & 0.1118 & 0.0010 & 0.1000 &-0.0455 &-0.0903 &-0.0347 &-0.0088 & 0.1256\\
5.5      & 0.0717 & 0.0948 & 0.1020 & 0.0012 & 0.0975 &-0.0617 &-0.0839 &-0.0468 &-0.0037 & 0.1222\\
6.0      & 0.0619 & 0.0858 & 0.0935 & 0.0014 & 0.0949 &-0.0776 &-0.0769 &-0.0583 & 0.0017 & 0.1185\\
6.5      & 0.0522 & 0.0772 & 0.0854 & 0.0018 & 0.0923 &-0.0936 &-0.0700 &-0.0695 & 0.0071 & 0.1146\\
7.0      & 0.0419 & 0.0687 & 0.0766 & 0.0021 & 0.0895 &-0.1095 &-0.0629 &-0.0808 & 0.0121 & 0.1106\\
\end{tabular}
\end{ruledtabular}
\label{tab:graphite:energychange}
\end{table*}

\textsl{Energy changes --}
Table~\ref{tab:graphite:energychange} shows the energy
changes calculated using Equation~\ref{eqn:deltaG} and \ref{eqn:deltaE}.
The positive free energy changes here are different
from the {\abin} study, \cite{chan.sp:2004}
mainly because of the different $p\Delta V$ values.
Our studies show a negative $p\Delta V$ when the pressure
is smaller than 5.8 GPa, in agreement with the {\abin} study.
However, the values are different.
For example, under 2.0 GPa the $p\Delta V$ term for
C(H$_2$)$_{1/2}$ of -0.0146 eV per H$_2$ molecule is much
smaller than the {\abin} result of -0.0975 eV.
The reason is easy to find out that in the {\abin} study,
the supercell contains only 8 H$_2$ molecules that even
under small pressure the hydrogen is fully confined.

We also provide the contributions to the total energy change
$\Delta E$ to analysis the storage mechanism.
One can find from the table that $\Delta E_{\mathrm{CC}}^1$ is
nearly zero.
This is because the graphene sheets keep their plane structure
all the time, without changing the covalent energy.
The positive $\Delta E_{\mathrm{CC}}^2$ terms show that the
storage process should overcome the binding energy between
graphene sheets.
Fortunately, the negative $E_{\mathrm{C-H}_2}$ is almost
in the same magnitude as $\Delta E_{\mathrm{CC}}^2$.
The cancellation between these terms makes only
$\Delta E_{\mathrm{H}_2-\mathrm{H}_2}$ left to the total energy change.
$\Delta E_{\mathrm{H}_2-\mathrm{H}_2}$ becomes more negative
with increasing the pressure
is because the hydrogen inside graphite is solid and can
not be further compressed along the two directions ($x$ and $y$)
of the graphene plane due to the graphene's hexagonal
structure, while the pure hydrogen can be compressed further
in all three directions.

In all three simulations, $\Delta E$ are negative beyond few GPa.
However, the magnitudes of $\Delta E$ and $p\Delta V$ are so
small that they can not overcome the quite large $T\Delta S$ term.
In other words, such layered structure under high pressure
is energetically favorable but not allowed due to the positive
free energy change.

\textsl{Pressure of hydrogen --}
We also calculate the pressure of hydrogen between the graphene sheets,
as shown in Figure~\ref{fig:graphite:h2_rho}.
Though the pressure of hydrogen inside is smaller than the external
pressure, more hydrogen intercalation is not allowed because of the
positive $\Delta G$.
Negative $\Delta G$ is the first rule for hydrogen storage, then
the pressure is another judgment that at the maximum storage the
pressure of hydrogen is no larger than the external one.

\section{Hydrogen inside carbon nanotubes}

\begin{figure}
\centering
 \includegraphics[width=3.6in]{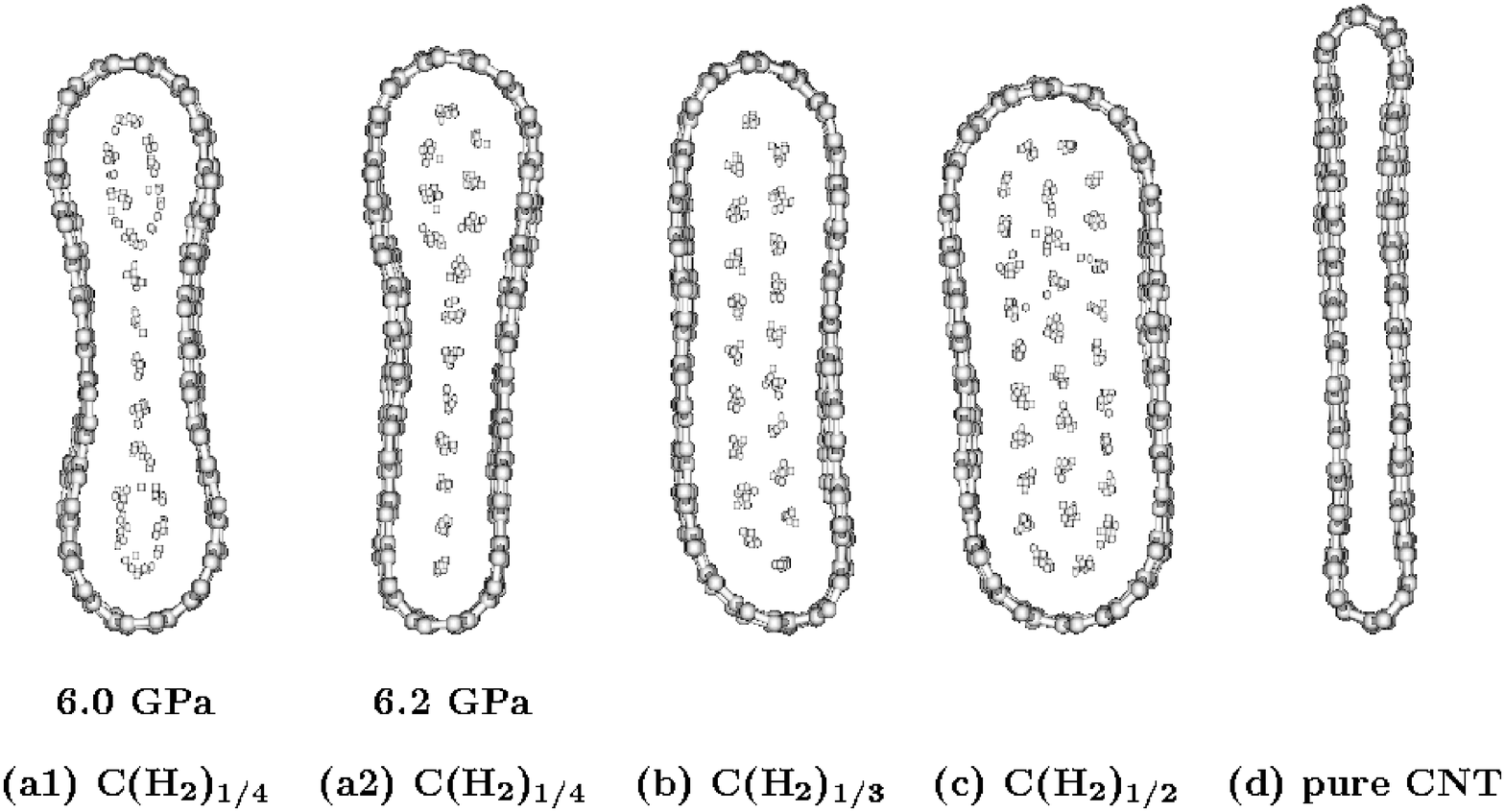}
\caption{Hydrogen structure inside (15,15) CNT under 6.0 GPa
at 300 K. Specially for the number ratio of C:H$_2$=4:1,
there is a phase transition under 6.1 GPa, with different
structures before (6.0 GPa) and after (6.2 GPa) the transition
shown in (a1) and (a2). In order to show the curvature effect,
the structure of pure CNT under 6.0 GPa is also shown.}
\label{fig:15_15:structure}
\end{figure}

\textsl{Hydrogen structure --}
Figure~\ref{fig:15_15:structure} shows the layered structure of
hydrogen inside the (15,15) CNT under high pressure.
The hydrogen-intercalated system also undergoes step-by-step
hydrogen condensations to reach the final layered hydrogen
structure, similar to hydrogen inside the graphite.
For the number ratio of C:H$_2$=4:1, there is another phase
transition under pressure about 6.1 GPa, that before the
transition only the middle part of the hydrogen is well confined
while after the transition the hydrogen show one-layered
structure in one part and two-layered in another.
These structures are shown in Figure~\ref{fig:15_15:structure}(a1)
and Figure~\ref{fig:15_15:structure}(a2), respectively.
For the $H2$ molecules, the main difference from the hydrogen
inside graphite is the molecules at the two ends of the ellipse
have more carbon neighbors and therefore have lower
energy than other H$_2$ molecules in the middle of the
ellipse.

On the other hand and more important, with hydrogen
intercalation, the CNT
wall relaxes itself to release its elastic energy.
One can find the different bent shapes for the tube
with and without hydrogen inside from Figure~\ref{fig:15_15:structure}.
For the pure carbon CNT under pressure, the shape of
the cross section is an elongated ellipse, with a large
elastic energy at the two ends of the ellipse.
But when the hydrogen is intercalated, the curvature
becomes larger at the two ends.
Such effect is new for CNT's, and will contribute
more negative energy change for hydrogen storage.

\begin{figure}
\centering
\includegraphics[height=3.2in,angle=-90]{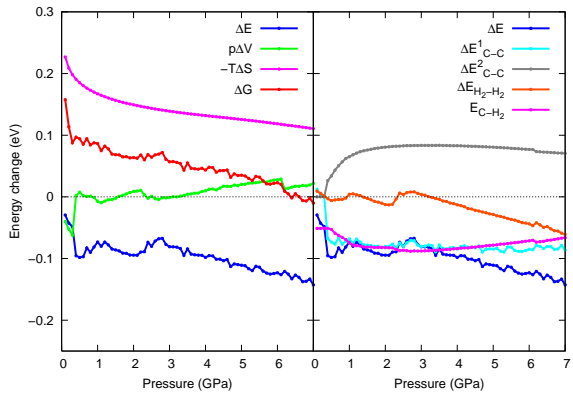}
\caption{Energy changes for (15,15) CNT: C(H$_2$)$_{1/4}$.
There are several jumps in the energy changes, corresponding
to different phase transitions or hydrogen condensations:
(i) the phase transition of pure (15,15) CNT (0.3 GPa),
(ii) the first hydrogen condensation (0.9-1.1 GPa),
(iii) the second hydrogen condensation (2.2-2.4 GPa),
and (iv) the final phase transition of hydrogen to
form a well confined structure (6.1 GPa), as described
in the text.}
\label{fig:15_15_1:energy}
\end{figure}

\textsl{Energy changes --}
As we expected, hydrogen storage inside CNT's produces
more negative $\Delta E$.
Fortunately, the magnitude of $p\Delta V$ is still quite
small despite of the sign, thus the free energy change
$\Delta G$ might be negative under pressure.
Figure~\ref{fig:15_15_1:energy} shows the energy changes
for hydrogen inside (15,15) CNT with the number ratio of
C:H$_2$=4:1.
\footnote{$\Delta E$ shows large fluctuation. However
it is acceptable, because for C-C covalent interaction,
the energy is around -7.35 eV per C atom through
current potential. The fluctuation is only 0.01 eV
per H$_2$ molecule, corresponding to 0.0025 eV per C atom,
about only 0.03\%.}
It is quite interesting to find that under pressure above 6.4 GPa,
$\Delta G$ is negative (-0.0016 and -0.0101 eV per H$_2$ molecule
for 6.4 and 7.0 GPa respectively).
For the number ratio of 3:1, $\Delta G$ is also negative under
7.0 GPa (-0.0011 eV).

Considering that in the $\Delta G$ calculation the entropy
difference between the pure and hydrogen intercalated carbon
system is taken to zero, the real $\Delta G$ should be a
little smaller, or in other words, more negative.
This result is quite different from the graphite that
under pressure in gigapascal range, hydrogen storage
inside carbon nanotube is free-energetically favorable.

When more and more hydrogen (C:H$_2$=2:1) is inside the tube,
we find $\Delta G$ increases to 0.0187 eV per H$_2$ molecule
even under 7.0 GPa.
It shows the maximum storage of hydrogen in our pressure
range is between the number ratio of C:H$_2$=3:1 and 2:1.

To find out the reason why hydrogen can be stored in CNT,
we still should look into the contributions to the total
energy change $\Delta E$.
In Figure~\ref{fig:15_15_1:energy} one can find that
the covalent bonding energy between carbon atoms contributes
now, shown in the $\Delta E_{\mathrm{CC}}^1$ term, as we
have explained above.
Hydrogen inside the tube relaxes the tube wall and cause
a negative energy change, which cancels the positive
vdW binding energy change $\Delta E_{\mathrm{CC}}^2$.
We will show that $\Delta E_{\mathrm{CC}}^1$ can be more
negative when the tube radius decreases.
However, for the left two contributions $\Delta E_{\mathrm{H}_2-\mathrm{H}_2}$
and $E_{\mathrm{C-H}_2}$, they are still as the same as
those in graphite.
In all, due to the curvature effect, hydrogen storage
inside CNT causes $\Delta E_{\mathrm{CC}}^1$ to be negative
and makes the total energy change $\Delta E$ large enough
to overcome the $T\Delta S$ term.

\begin{figure}
\centering
\includegraphics[height=3.2in,angle=-90]{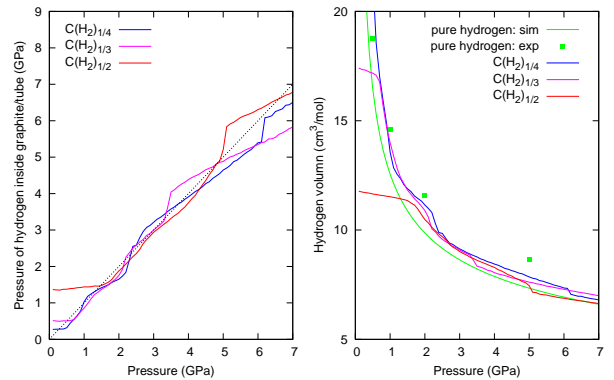}
\caption{Hydrogen pressure for the storage inside (15,15) CNT
along the direction ($z$) of the tube axis (left) and
the volume of the hydrogen, similar to Figure~\ref{fig:graphite:h2_rho}.}
\label{fig:15_15_1:rho}
\end{figure}

\textsl{Pressure of hydrogen --}
In Figure~\ref{fig:15_15_1:rho} we show the calculation
of the hydrogen pressure along the tube axis direction ($z$)
and the volume for the hydrogen.
We can also get the maximum storage capacity for (15,15) CNT
from the figure.
For the number ratio of C:H$_2$=4:1 and 3:1, $\Delta G$ is
negative under $\sim$ 7.0 GPa.
At the same time, the pressure inside is smaller than the
external one, thus more hydrogen can be stored due to
the negative free energy change.

\begin{table}[b]
\caption{$\Delta G$ for (10,10) and (15,15) CNT.
$x$ is the number ratio between molecular hydrogen and carbon
atoms, labeled by C(H$_2$)$_x$. The unit for $\Delta G$ is
eV per H$_2$ molecule.}
\begin{ruledtabular}
\begin{tabular}{ccccccc}
Pressure & \multicolumn{3}{c}{$\Delta G$ for (10,10) CNT}
         & \multicolumn{3}{c}{$\Delta G$ for (15,15) CNT}\\
\cline{2-4}\cline{5-7}
(GPa)    & $x=1/5$ & 1/4 & 3/8 & $x=1/4$ & 1/3 & 1/2 \\
\hline
0.6      & 0.0164&       &       &       &       &       \\
0.7      &-0.0063& 0.0209&       &       &       &       \\
0.8      &-0.0204& 0.0676&       &       &       &       \\
0.9      &-0.0414&-0.0048&       &       &       &       \\
1.0      &       &-0.0194&       &       &       &       \\[2ex]
3.8      &       &       & 0.0006&       &       &       \\
3.9      &       &       & 0.0020&       &       &       \\
4.0      &       &       &-0.0001&       &       &       \\
4.1      &       &       &-0.0058&       &       &       \\[2ex]
6.3      &       &       &       & 0.0050&       &       \\
6.4      &       &       &       &-0.0016&       &       \\
6.5      &       &       &       & 0.0051&       &       \\
6.6      &       &       &       & 0.0004&       &       \\
6.7      &       &       &       &-0.0060& 0.0005& 0.0211\\
6.8      &       &       &       &-0.0067& 0.0009& 0.0204\\
6.9      &       &       &       &-0.0026& 0.0022& 0.0220\\
7.0      &       &       &       &-0.0101&-0.0011& 0.0187\\
\end{tabular}
\end{ruledtabular}
\label{tab:cnt:energychange}
\end{table}

\textsl{Smaller tube --}
When the radius of CNT decreases, the curvature plays more important
role to the energy change.
In our simulations for (10,10) CNT, we find that $\Delta G$ is
negative even under $<$1.0 GPa, much smaller than the critical
pressure for (15,15) CNT.
Table~\ref{tab:cnt:energychange} also shows that the critical
pressure for a negative $\Delta G$ increases with increasing
the number of hydrogen.

\section{Conclusion}

In summary, we have studied the hydrogen storage inside
graphite and carbon nanotubes.
Three results are listed below:

i) Hydrogen can not be stored inside graphite due
to the positive energy change.
The main reason is that the binding energy between
graphene sheets is overcome only by the C-H$_2$
vdW interaction and the H$_2$-H$_2$ energy change.
So the total energy change $\Delta E$ is not
large enough to overcome the positive $T\Delta S$ term.
The hydrogen inside graphite is solid with hcp
structure, and has a smaller internal pressure
than the external one.

2) For the carbon nanotube, the hydrogen storage
under pressure will relax the tube wall and bring
an additional contribution to the total energy change.
So it is possible to achieve a negative free energy change.
For example, under $\sim$ 7.0 GPa, hydrogen can be
stored inside (15,15) CNT due to the negative free
energy change, with the storage capacity higher
than 5.2 wt\%.

3) When the tube radius decreases, the curvature
effect plays more important role in the total energy change.
The negative free energy change can be achieved
even under a small pressure.

Our results also show the possibility to design the
CNT-based nanocontainer. \cite{ye.x:2007}
The hydrogen can be filled inside the tube under
pressure in gigapascal range,
and because of the molecular valves, when we
unload the external pressure, molecular hydrogen
will be locked inside the tube.
Such fill-and-lock mechanism has been discussed
by Ye \etal, \cite{ye.x:2007}
and is proved applicable by current studies.

\begin{acknowledgements}
This work is supported by an Earmarked Grant
(Project No. CUHK 402305P) from the Research
Grants Council of Hong Kong SAR Government,
and partially by the National Science Foundation
of China, the special funds for major state basic
research.
The computation is performed in the Supercomputer
Center of Fudan University.
\end{acknowledgements}

\bibliography{reference}

\end{document}